\documentclass[12pt]{iopart}
\usepackage{graphicx}

\begin{document}

\title[Y. G. Ma]{Experimental observables on nuclear  liquid gas phase transition \footnote[3]{
Dedicate to the "International Seminar on Atomic and Nuclear
Cross-Disciplinary Physics" in honor of Prof. Fujia Yang's 70th
Birthday. }}

\author{Y. G. Ma  }

\address{ Shanghai Institute of Applied Physics, Chinese Academy of Sciences,  China\\ E-mail: ygma@sinap.ac.cn}

\begin{abstract}
Progress on nuclear liquid gas phase transition (LGPT)  or
critical behavior has been simply reviewed and some signals of
LGPT in heavy ion collisions, especially in NIMROD data, are
focused. These signals include the power-law charge distribution,
the largest fluctuation of the fragment observables, the nuclear
Zipf law, caloric curve and critical exponent analysis etc.
\end{abstract}



%

\section{Introduction}

Phase transition and critical phenomenon are extensively debatable
subjects in the natural sciences. Recently, the same concept was
introduced into the astronomical objects \cite{Lynden1} as well as
the microscopic systems, such as in atomic cluster \cite{Bertsch2}
and nuclei \cite{Purdue1}. Nucleus is a finite-size droplet-like
quantum system in the ground state, it may experience  the
liquid-gas phase transition or critical behavior when the nucleus
is heated. Till date, various experimental evidences have been
accumulated to corroborate the nuclear liquid gas phase transition
or critical behavior. For instance, violent heavy-ion collisions
break the nuclei into several intermediate mass fragments (IMF)
instantaneously \cite{Peaslee94,Ogilvie91,Tsang93,Ma_PRC95}, which
might be viewed as a critical phenomenon as observed in fluid,
atomic and other systems. The sudden opening of the nuclear
multifragmentation and vaporization \cite{Rivet4} channels can be
interpreted as the signature of the boundaries of phase mixture.
In addition, temperature  plateau of the nuclear caloric curve
\cite{Gross5} in a certain excitation energy range gives a
possible indication of phase transition
\cite{Poch6,JBN7,Ma_PLB97,JBN_PRL}. On the other hand, the
extraction of critical exponents in the charge or mass
distribution of the multifragmentation system \cite{EOS_expo}
points to the continuous phase transition. More recently, the
negative microcanonical heat capacity was experimentally attempted
in nuclear fragmentation \cite{Gul-WCI} which may relate to the
liquid-gas phase transition \cite{Chomaz11}, and in atomic cluster
\cite{Schm12} which relates to solid-liquid phase transition
\cite{Lab13}, respectively. Moreover, the evidence of spinodal
decomposition in nuclear multifragmentation was recently reported
experimentally \cite{Bord14,Borderie-WCI}, which may show the
presence of liquid-gas phase coexistence region in the finite
nuclear systems. Bimodal distributions of some chosen variables
measured in nuclear collisions were recently proposed as a
signature of a first order phase transition in nuclei
\cite{Lopez-WCI}. $\Delta$-scaling of the largest fragments was
also investigated recently and it is probably  a probe to detect
the phase change \cite{Botet_PRL01}. The nuclear Zipf's law was
proposed to diagnose the onset of liquid-gas phase transition or
critical behavior \cite{Ma_PRL99,Ma_CPL,Ma-review} and it is
experimentally supported \cite{Ma2005,Dabrowska}. Phase
coexistence diagram was also tentatively constructed based on the
EOS data \cite{elliott1,Ellio03}.

While, extensive theoretical models have  been developed to treat
such a phase transition in the nuclear disassembly, e.g.
percolation model, lattice gas model, statistical
multifragmentation model and molecular dynamics model etc (e.g.
see some recent articles
\cite{Rich15,Das_rev,Bonasera,Chomaz_INPC,Moretto_INPC,Ma_rev} and
references therein).

In this article which dedicates to Prof Fu-Jia Yang on the
occasion of his 70th birthday, I will focus on some experimental
aspects of the nuclear LGPT or critical behavior in heavy ion
collisions, specially for NIMROD data.

\section{Signals of the Nuclear Liquid Gas Phase Transition or Critical Behavior}

\subsection{Fisher Droplet Model: Power Law of Charge/Mass Distribution}

The Fisher droplet model has been extensively applied to the
analysis of multifragmentation since  the pioneering experiments
on high energy proton-nucleus collisions by the Purdue group
\cite{Purdue1,Purdue2,Minich82}. Relative yields of fragments with
$3 \leq Z \leq 14$ could be well described by a power law
dependence $A^{-\tau}$ and it was suggested that this might
reflect the disassembly of a system whose excitation energy was
comparable to its total binding energy \cite{Purdue2}. The
extracted value of power-law exponent was $2 \leq \tau \leq 3$,
which is in a reasonable  range for critical behavior
\cite{Fisher}. The success of this approach suggested that the
multi-fragmentation of nuclei might be analogous to a continuous
phase transition from liquid to gas which is observed in more
common fluids.

In the Fisher Droplet Model  the fragment mass distribution can be
represented as
\begin{equation}
Y(A) = Y_0 A^{-\tau} X^{A^{2/3}}Y^A,
\end{equation}
where $Y_0$, $\tau$, $X$ and $Y$ are parameters. However, at the
critical point $X$ = 1 and $Y$ = 1, therefore the cluster
distribution is given by a pure power law
\begin{equation}
Y(A) = Y_0 A^{-\tau}.
\end{equation}
The model predicts a critical exponent $\tau \sim$ 2.21.

\begin{figure}
\includegraphics[scale=0.4]{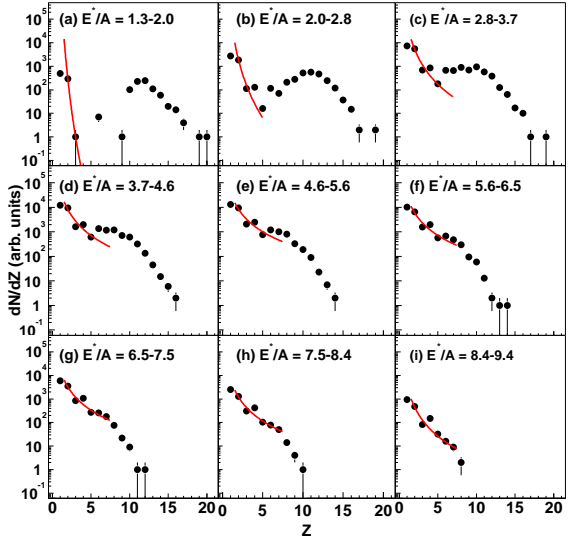}
\caption{\footnotesize (Left Part) Charge distribution of QP in
different $E^*/A$ (in unit of MeV/nucleon) window. From the TAMU
data \cite{Ma2005}. } \label{fig_Zdist}
\end{figure}

\begin{figure}
\vspace{-4.4truein} \hspace{3.3truein}
\includegraphics[scale=0.4]{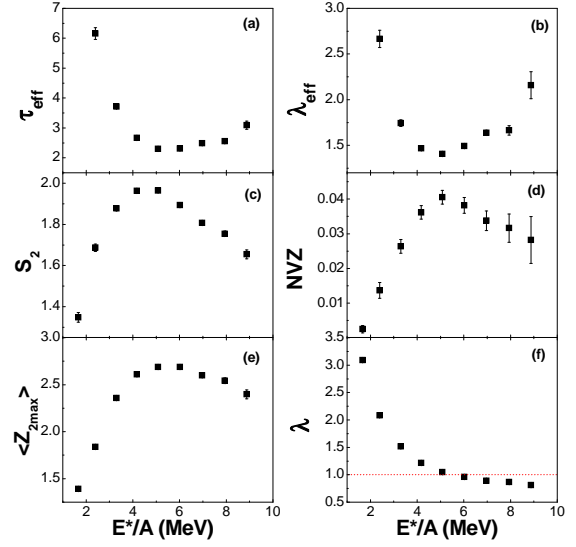}
\vspace{-1.cm} \caption{(Right part) The effective Fisher-law
parameter ($\tau_{eff}$) (a), the effective exponential law
parameter ($\lambda_{eff}$) (b), $\langle S_2\rangle$ (c), NVZ
fluctuation (d),   the mean charge number of the second largest
fragment $\langle Z_{2max}\rangle$ (e), the Zipf-law parameter
$\lambda$ (f). See details in text.  Figure is taken from the TAMU
data \cite{Ma_NPA2005}.}
 \label{fluctuation}
\end{figure}

We collaborated an   experiment with a 4$\pi$-detector combination
of charged particle ball with neutron ball (NIMROD) in Cyclotron
Institute, Texas A\&M University to explore the quasi-projectile
fragmentation from Ar + Al, Ti and Ni at 47 MeV/nucleon
\cite{Ma2005}. Well defined QP sources have been reconstructed
using a new method which is based on three source fits and the
Monte Carlo sampling for the assignment of quasi-projectile (QP)
and their disassembly features have been analyzed. For simplicity,
we call this experimental data  as TAMU data \cite{Ma2005}. The
details can be found in Ref.~\cite{Ma2005}.

Fig.~\ref{fig_Zdist} presents  yield distributions  $dN/dZ$,
observed for nine different excitation energy ($E^*/A$) intervals
for the QP from the reactions of $^{40}Ar$ + $^{58}Ni$ from TAMU
data. At low  excitation energy a large $Z$ residue always
remains, $i.e.$ the nucleus is basically in the liquid phase
accompanied by some evaporated light particles. When $E^*/A$
reaches  $\sim$ 5.6 MeV/nucleon, this residue is much less
prominent.  As $E^*/A$ continues to increase, the charge
distributions become steeper which indicates that the system tends
to vaporize. To quantitatively pin down the possible phase
transition point, we use a power law fit to the QP charge
distribution in the range of $Z$ = 2 - 7  to extract the effective
Fisher-law parameter $\tau_{eff}$ by $ dN/dZ \sim Z^{-\tau_{eff}}$
in Fig.~\ref{fig_Zdist}.  Fig.~\ref{fluctuation}(a) shows
$\tau_{eff}$ vs excitation energy, a minimum with $\tau_{eff}$
$\sim$ 2.3 is seen in the $E^*/A$ range of  5 to 6 MeV/nucleon.
$\tau_{eff}$ $\sim$ 2.3, is close to the  critical exponent of the
liquid gas phase transition universality class as predicted by the
Fisher's droplet model \cite{Fisher}.

While, assuming that the heaviest cluster in each event represents
the liquid phase, we have attempted to isolate the gas phase by
event-by-event removal of the heaviest cluster from the charge
distributions. We found that the resultant distributions are
better described as exponential form $ e^{-\lambda_{eff} Z}$. The
fitting parameter $\lambda_{eff}$  was derived and was plotted
against excitation energy in Fig.~\ref{fluctuation}(b). A minimum
is seen in the same region where $\tau_{eff}$ shows a minimum.

\subsection{The largest fluctuation}

One of the well known characteristics of the systems undergoing a
continuous phase transition is the occurrence  of the largest
fluctuations. These large fluctuations in cluster size and density
of the system arise because of the disappearance of the latent
heat at the critical point. In macroscopic systems such behavior
gives rise to the phenomenon of critical opalescence
\cite{Stanley}.

To further explore this region we have investigated other proposed
observables commonly related to fluctuations and critical
behavior. Fig.~\ref{fluctuation}(c) shows the excitation function
of the mean normalized second moment, $\langle S_2\rangle$,
defined as $S_2 = {M_2}/{M_1}$, where $M_2$ and $M_1$ is the
second and first moment. Generally, the $k$-th moment is defined
as
\begin{equation}
M_k= \sum_{A \ne A_{max}} A^k m, \label{eqn-Mk}
\end{equation}
where $m$ is the multiplicity of the fragment $A$ and the largest
fragment $A_{max}$ is excluded in the summer
\cite{Campi,Ma-review}. A peak is seen around 5.6 MeV/nucleon, it
indicates that the fluctuation of the fragment distribution is the
largest in this excitation energy region. Similarly,  the
Normalized Variance in $Z_{max}/Z_{QP}$ distribution (i.e. NVZ =
$\frac{\sigma^2_{Z_{max}/Z_{QP}}}{\langle Z_{max}/Z_{QP}\rangle}$)
\cite{Dorso} shows a maximum in the same excitation energy region
[Fig.~\ref{fluctuation}(d)], which illustrates the maximal
fluctuation for the largest fragment is reached around $E^*/A$ =
5.6 MeV. Except the largest fragment, the second largest fragment
also shows its importance in the above turning point.
Fig.~\ref{fluctuation}(e) shows a broad peak of $\langle
Z_{2max}\rangle$ - the average atomic number of the second largest
fragment exists at 5.6 MeV/nucleon.

\subsection{Fragment Hierarchical Distribution: Zipf plot and Zipf law }

In addition to the largest fluctuation, observables revealing some
particular topological structure may also reflect the critical
behavior for a finite system. In this section we discuss a new
observable and law: nuclear Zipf plot and Zipf law.

Zipf's law has been known as a statistical phenomenon concerning
the relation between English words and their frequency in
literature in the field of linguistics \cite{Zipf}. The law states
that, when we list the words in the order of decreasing
population, the frequency of a word is inversely proportional to
its rank \cite{Zipf}. Recently, Ma proposed  measurements of the
fragment hierarchy distribution as a tool to search for the liquid
gas phase transition in a finite nuclear system based on the
calculations with isospin-dependent lattice gas model (I-LGM) and
classical molecular dynamics model \cite{Ma_PRL99,Ma_EPJA}. The
fragment hierarchy distribution can be defined by the so-called
Zipf plot, i.e., a plot of the relationship between mean sizes of
fragments which are rank-ordered in size, i.e., largest, second
largest, etc. and their rank \cite{Ma_PRL99}.

\begin{figure}
\hspace{1.5cm} \vspace{-0.1truein}
\includegraphics[scale=0.6]{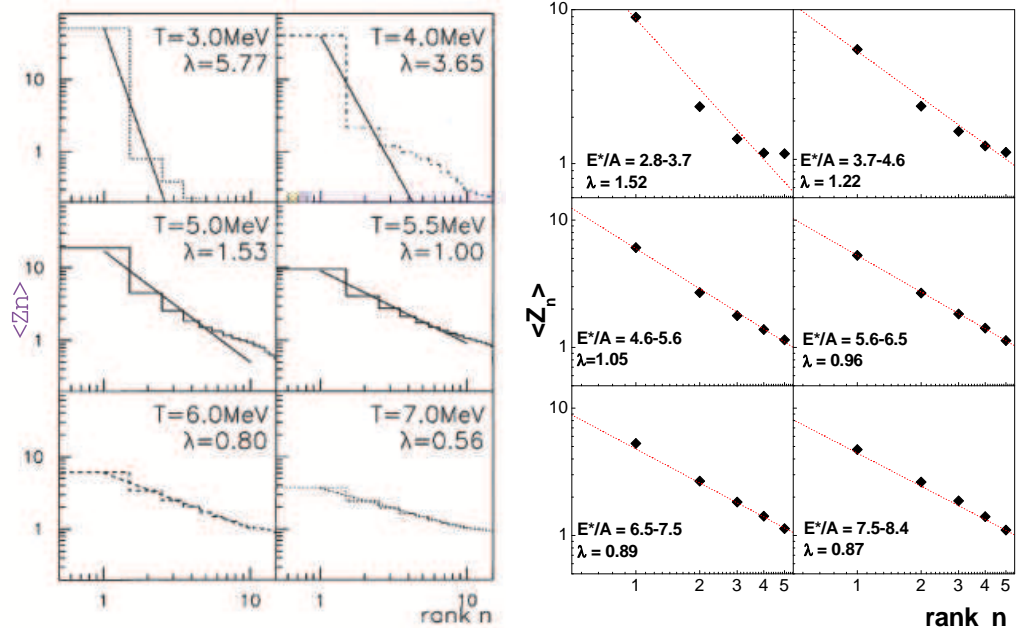}
\vspace{0.2truein} \caption{\footnotesize (Left part) Zipf plots
for the fragments from $^{129}Xe$ at temperature $T$ = 3 - 7 MeV
with I-LGM calculation. The histogram are simulation rseults and
the lines are Zipf power-law fits (Eq.~\ref{eq_zipf}). From Ref.
\cite{Ma_EPJA}.} \label{fig_epj_zipf}
\end{figure}

\begin{figure}
 \vspace{-4.9truein}
 \hspace{8.5cm}
   \includegraphics[scale=0.37]{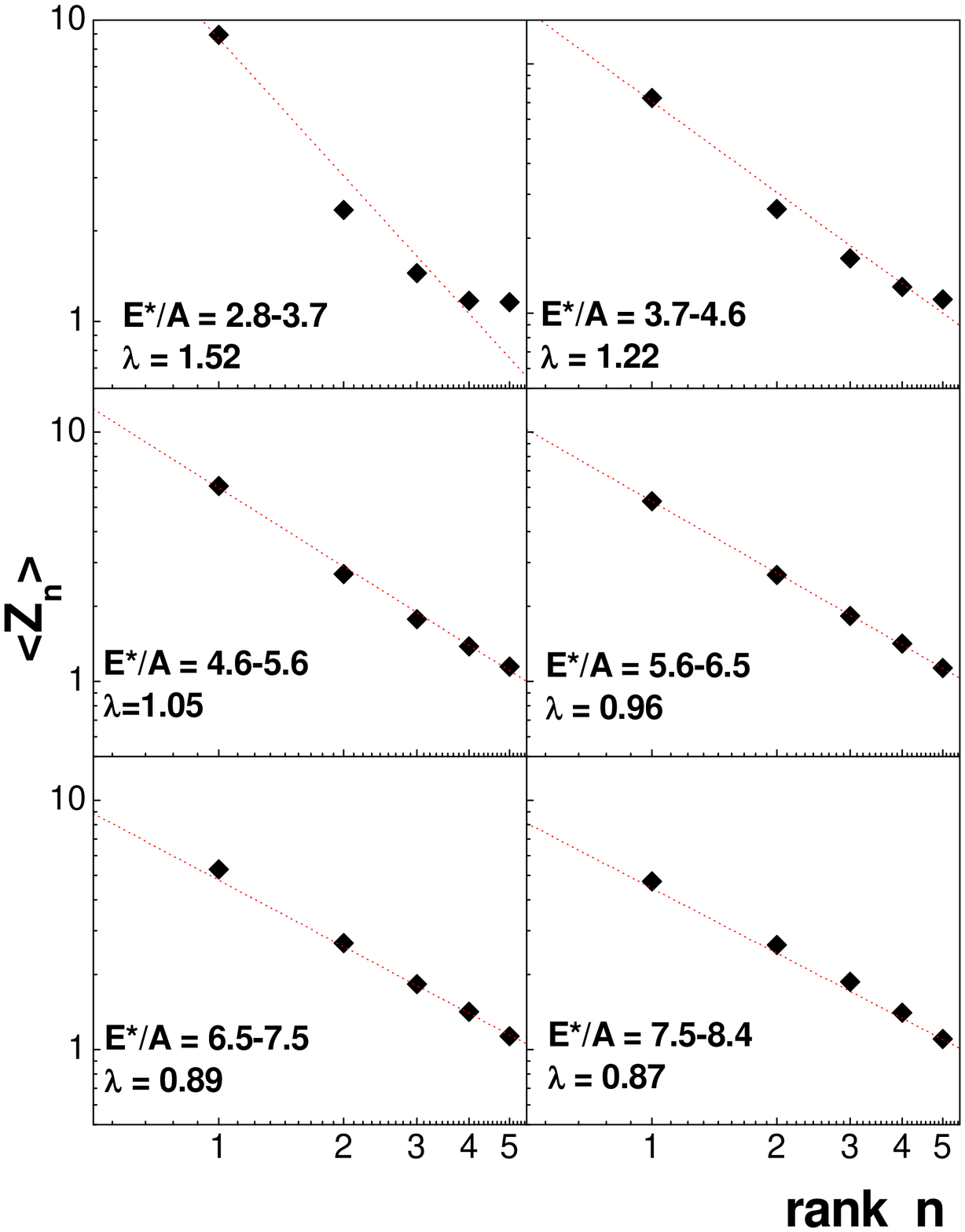}
\vspace{0.2truein} \caption{\footnotesize (Right part) Zipf plots
in six different excitation energy bins for   the QP formed in
$^{40}$Ar + $^{58}$Ni. The dots are data and the lines are
Zipf-law fits. The statistical error is smaller than the size of
the symbols.  From Ref. \cite{Ma-review}. } \label{fig_zipf}
\end{figure}

In such an analysis, the cluster size is employed  as a variable
to make a Zipf-type plot, and the resultant distributions can be
fitted with a  power law,
\begin{equation}
\langle Z_{n}\rangle \propto  n^{-\lambda}, \label{eq_zipf}
\end{equation}
where $n$ represents the rank and  $\lambda$ is the Zipf's law
parameter. Fig.~\ref{fig_epj_zipf} shows the Zipf-type plots for
the disassembly of $^{129}Xe$ with I-LGM calculations
\cite{Ma_EPJA}. The fit parameter $\lambda$ was extracted in each
temperature. When $\lambda \sim 1$,  Zipf's law is satisfied.  In
this case, the mean size of the second largest fragment is 1/2 of
that of the the largest fragment;  That of the third largest
fragment is 1/3 of the largest fragment, etc.

Now we return to the TAMU data \cite{Ma2005}, as shown in
Fig.~\ref{fig_Zdist}, the region of 5-6 MeV/nucleon excitation
energy is related to the critical behavior as shown in above
sections. The significance of this region in our data is further
indicated by a Zipf's law analysis. Fig.~\ref{fig_zipf} shows the
Zipf plots in six excitation energy windows. This rank ordering of
the probability observation of fragments size of a given rank,
from  the largest to the smallest, does indeed lead to a Zipf's
law power law parameter $\lambda$ $\sim$ 1 in the 5-6 MeV/nucleon
range.

In a recent analysis for  multifragment emission for CERN EMU13
data, Dabrowska et al. analyzed emulsion data for 158 A GeV Pb-Pb
and Pb-Plastic collisions and tested the nuclear Zipf law of Ma
\cite{Dabrowska}. They found that their data are roughly
consistent with the nuclear Zipf law in a certain multiplicity
where  the multiplicity of  IMF and $\tau$ parameter of the charge
distribution reveal the same turning point. This has been
interpretated as an evidence for the existence of the critical
temperature associated with a liquid gas phase transition
\cite{Dabrowska}.

\subsection{Caloric Curve }

The caloric curve which relates the internal energy of an excited
system at thermodynamic equilibrium to its temperature is a priori
the simplest experimental tool to look for the existence of a
phase transition.

In our TAMU data the caloric curve was constructed \cite{Ma2005}
and a monotonous rising behavior was observed instead of a
plateau-like structure observed in heavier nuclei \cite{JBN7}. The
caloric curve of this lighter system looks more like cross over
behavior rather than first order phase transition. The
corresponding initial temperature is determined and it is  8.3
$\pm$ 0.5 MeV in the critical region of 5.6 MeV/nucleon excitation
energy \cite{Ma2005}.

\begin{figure}
\vspace{-0.8truein} \hspace{1.5cm}
\includegraphics[scale=0.4]{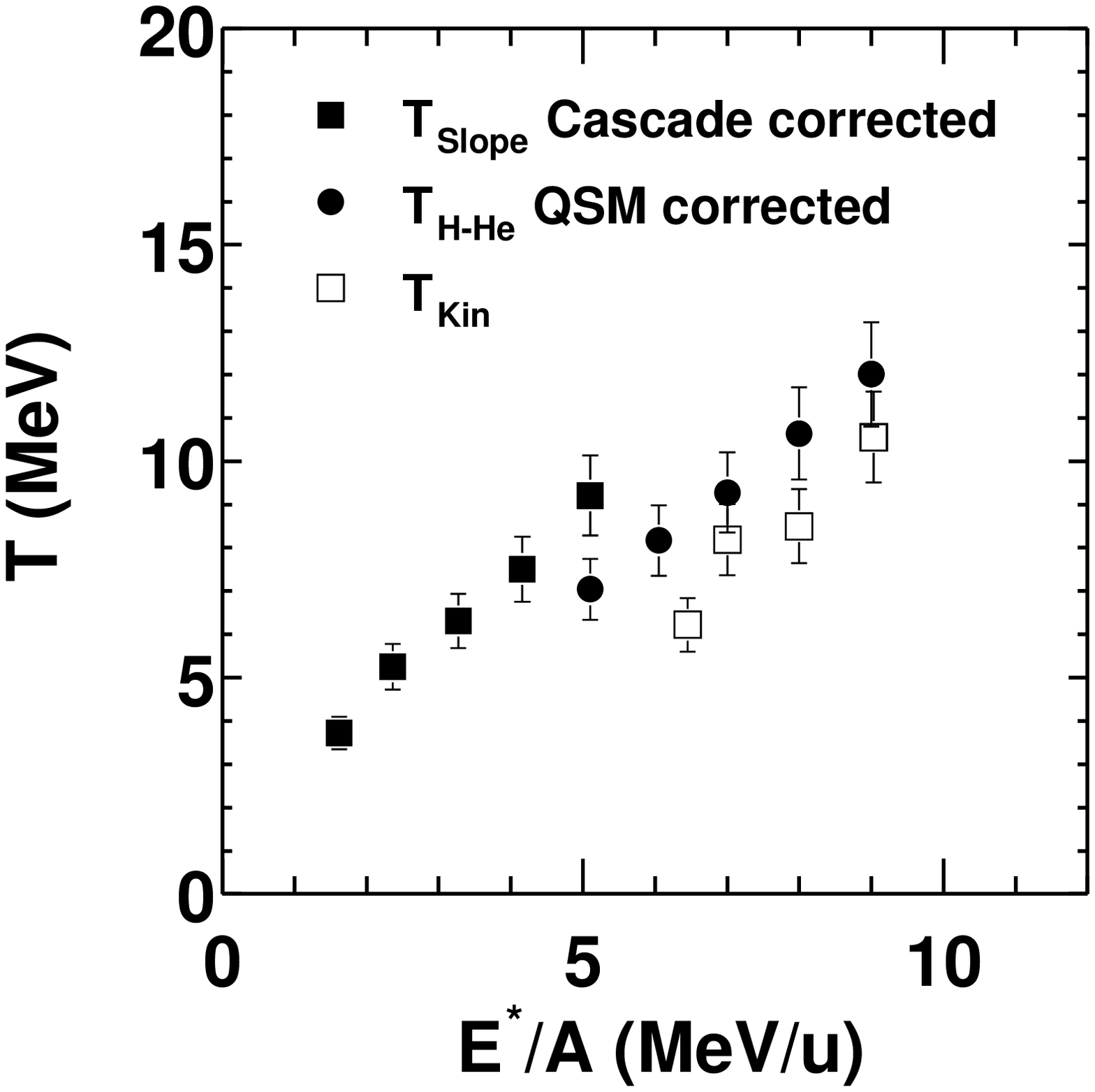}
\vspace{0.2truein} \caption{\footnotesize (Left part) The deduced
caloric curves for Ar quasi-projectiles ($A\sim 36$) in TAMU data.
The meanings of symbols are displayed in insert. For details, see
Ref. \cite{Ma2005}.} \label{fig_caloric}
\end{figure}

\begin{figure}
\vspace{-4.5truein} \hspace{9.0cm}
\includegraphics[scale=0.35]{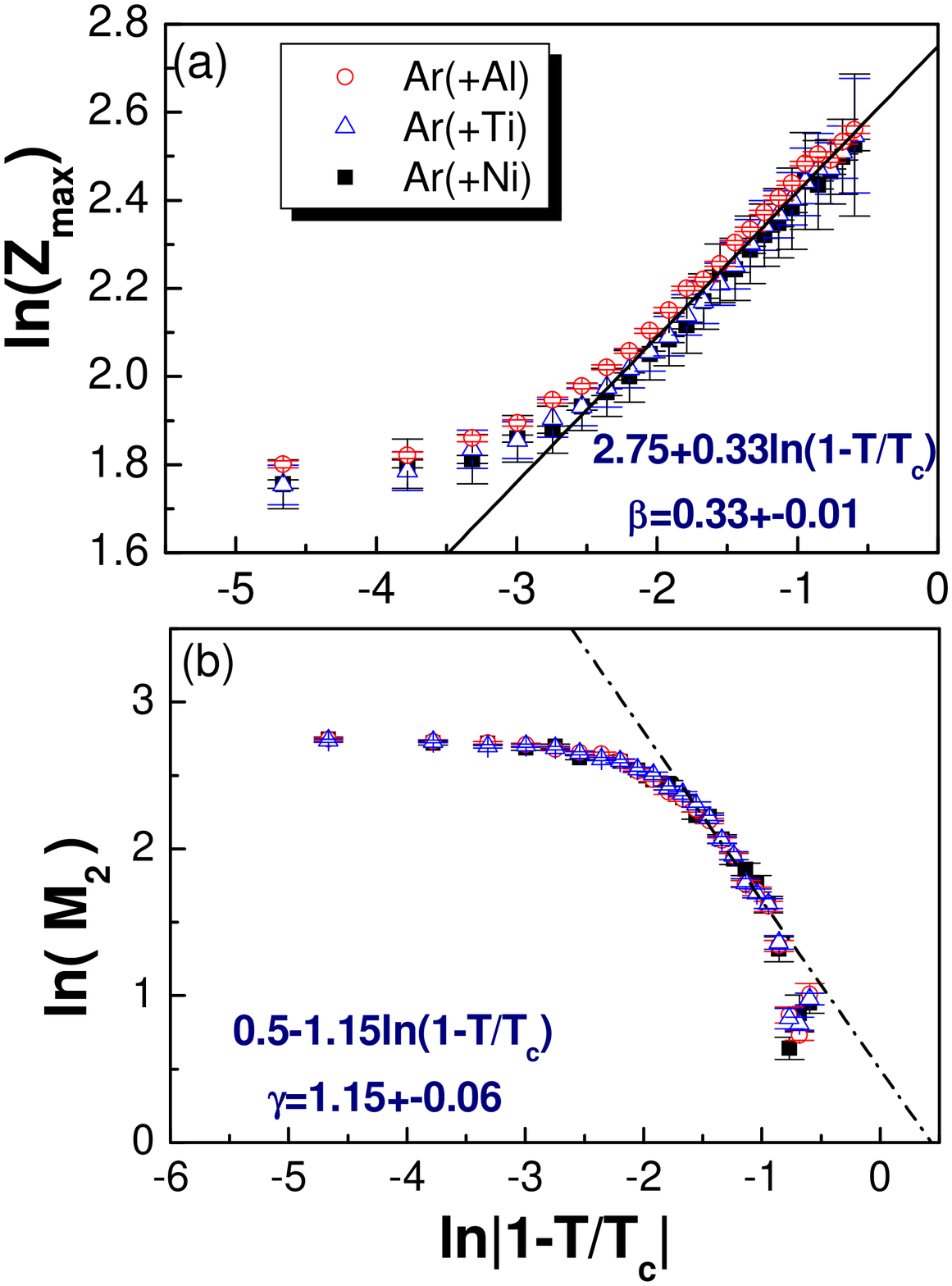}
\vspace{0.1truein}  \caption{\footnotesize (Right part) The
extraction of the critical exponent $\beta$ (a) and $\gamma$ (b).
See texts for details. } \label{fig_beta-gamma}
\end{figure}

\subsection{Critical Exponents}

 In terms of the scaling theory,
some critical exponents can also be extracted from the moment
analysis of cluster distributions \cite{Campi,Ma-review}, namely
$M_k$ of Eq.~\ref{eqn-Mk}. Since, for our system, we have already
deduced a critical temperature $T_c$ = 8.3 MeV at point of maximal
fluctuations \cite{Ma2005} from the caloric curve, we can use
temperature as a control parameter to test the critical exponents.
The critical exponent $\beta$ can be extracted from the relation
\begin{equation}
Z_{max} \propto (1-\frac{T}{T_c})^\beta, \label{eq_beta}
\end{equation}
and the critical exponent $\gamma$ can be extracted from the
second moment, via
\begin{equation}
M_2 \propto |1-\frac{T}{T_c}|^{-\gamma}. \label{eq_gamma}
\end{equation}
In each, $|1-\frac{T}{T_c}|$ is the parameter which measures the
distance from the critical point.

The upper panel of Fig.~\ref{fig_beta-gamma}  explores the
dependence of $Z_{max}$ on $(1- \frac{T}{T_c})$. A dramatic change
of $Z_{max}$ around the critical temperature  $T_c$ is observed.
LGM calculations also predict that the slope of $Z_{max}$ vs $T$
will change at the liquid gas phase transition \cite{Ma_JPG01}.
Using the liquid side side points, we can deduce the critical
exponent $\beta$ by $ln(Z_{max})$ vs $ln|1-T/T_c|$.
  An excellent fit was obtained in the region
away from the critical point and a critical exponent $\beta$ =
0.33 $\pm$ 0.01. Near  the critical point, the finite size effects
become stronger so that the scaling law is violated.

To extract the critical exponent $\gamma$, we take $M_2$ on the
liquid side without $Z_{max}$. Fig.~\ref{fig_beta-gamma}(b) shows
ln($M_2$) as a function of ln($|1-\frac{T}{T_c}|$). We center our
fit to Eq.~(\ref{eq_gamma}) about the center of the range of
$(1-T/T_c)$ which leads to the linear fit and extraction of
$\beta$ as represented in Figure.~\ref{fig_beta-gamma}(b). The
critical exponent $\gamma$ = 1.15 $\pm$ 0.06. It is seen that the
selected region has a good power law dependence.

Since the critical exponents $\beta$ and $\gamma$ have been
obtained, the critical exponent $\sigma$ can be deduced by the
scaling relation $\sigma = \frac{1}{\beta + \gamma}.$ In such way,
$\sigma$ = 0.68 $\pm$ 0.04 is obtained, which is also very close
to the expected critical exponent of a liquid gas system.

To summarize critical exponents of  our NIMROD data, we present
the results in Table 1 as well as the values expected for the 3D
percolation and liquid gas system universality classes. Obviously,
our values for this light system with A$\sim$36 are consistent
with the values of the liquid gas phase transition universality
class rather the 3D percolation class.

\begin{table}
\caption{Comparison of the Critical Exponents} \label{tab:1}
\begin{tabular}{llll}
\hline\noalign{\smallskip}
Exponents & 3D Percolation & Liquid-Gas & Our Data \\
\noalign{\smallskip}\hline\noalign{\smallskip}
$\tau$ & 2.18 & 2.21 & 2.31$\pm$0.03 (Fig.1)\\
$\beta$ & 0.41 & 0.33 & 0.33$\pm$0.01 \\
$\gamma$ & 1.8 & 1.23 & 1.15$\pm$0.06 \\
$\sigma$ & 0.45 & 0.64 & 0.68$\pm$0.04 \\
\noalign{\smallskip}\hline
\end{tabular}
\end{table}

\section{Conclusions }

In conclusion, the TAMU NIMROD data was focused and some
observables relating to the liquid gas phase transition or
critical behavior have been presented.  Around 5.6 MeV/nucleon of
excitation energy for A$\sim$36 system,  the fragment yield
distribution obeys the power law which was predicted by Fisher
droplet model. From fragment size structure, we found that there
exists a particular hierarchical arrangement, so-called the
nuclear Zipf law, which was demonstrated by both  Texas A\&M
 heavy ion data and EMU13 CERN emulsion data.
While, many observables demonstrate the existence of maximal
fluctuations around this excitation energy. These fluctuation
observables include the Campi scattering plots and the normalized
variances of the distributions of the largest fragment
($Z_{max}$). Caloric curve was reconstructed, and then critical
temperature and critical exponents are also discussed. Caloric
curve shows a gradual rise of temperature versus excitation
energy, no plateau is observed for such lighter system.  Thanks to
the determination of critical temperature from the caloric curve,
critical exponents are deduced and the values are close to the
liquid gas universal class.

Finally, we should mention that a unique signal can not give any
definite information whether the system is in a critical point or
is undergoing a phase transition since we are facing a transient
finite charged  system. Only many coherent signals, such as the
fluctuation peak, critical exponents, Fisher scaling as well as
Zipf's law etc, emerging together can corroborate the observation
of a phase transition or a critical behavior in finite nuclei.

\vspace{0.5cm} Author appreciates many colleagues and/or friends
for  discussions and collaborations, especially thanks Profs. F.
J. Yang, W. Q. Shen, J. B. Natowitz, J. P\'eter, B. Tamain, S. Das
Gupta, J. C. Pan et al. The paper is a contribution to a workshop
dedicated to Professor Fu-Jia Yang, former President of Fudan
University, former Director of Shanghai Institute of Nuclear
Research (now Shanghai Institute of Applied Physics), and the
Chancellor of Nottingham University, on his 70th birthday.
Professor Yang has been an inspiration for me as a scientist since
I got to know him when I was  a PhD student in his Institute. He
always  educates and encourages me how to become a good scientist,
and always gives his unselfish invaluable helps in my professional
life. I'd like to take this opportunity to make a grateful
acknowledgement for Professor Yang. This work was supported in
part by the National Natural Science Foundation of China (NNSFC)
for the Distinguished Young Scholar under Grant No 19725521 and
NNSFC under  Grant Nos. 10535010, 19705012, 10328259 and 10135030,
the Shanghai Development Foundation for Science and Technology
under Grant Numbers 06JC14082, 05XD14021 and 97QA14038, the Major
State Basic Research Development Program under Contract No
G200077404.

{}
\end{document}